\documentclass[12pt, a4paper]{article}
\usepackage{amsmath, amssymb, amsthm}
\usepackage{mathtools}
\usepackage{graphics}
\usepackage{graphicx}
\usepackage{epsfig}
\usepackage{dblfloatfix}
\usepackage{float}
\usepackage{placeins}
\usepackage{flafter}
\usepackage[usenames]{color}
\newtheorem{theorem}{\bf Theorem}

\newtheorem{proposition}{\bf Proposition}
\newtheorem{example}{\bf Example}

\usepackage{epstopdf}
\usepackage{url}
\usepackage{hyperref}
\date{}
\usepackage{fullpage}
\usepackage{setspace}

\newtheorem{df}{Definition}[section]

\newtheorem{cor}{Corollary}[section]

\makeatletter

\renewcommand\section{\@startsection {section}{1}{\z@}
{-30pt \@plus -1ex \@minus -.2ex} {2.3ex \@plus.2ex}
{\normalfont\normalsize\bfseries}}

\renewcommand\subsection{\@startsection{subsection}{2}{\z@}
{-3.25ex\@plus -1ex \@minus -.2ex} {1.5ex \@plus .2ex}
{\normalfont\normalsize\bfseries}}

\renewcommand{\@seccntformat}[1]{\csname the#1\endcsname. }

\makeatother
\title{\bf{Skew cyclic codes over $F_{p}+uF_{p}+\dots +u^{k-1}F_{p}$}}

\author{ \bf Om Prakash and Habibul Islam \\\\
Department of Mathematics \\
Indian Institute of Technology Patna\\ Patna- 801 106, India \\
E-mail: om@iitp.ac.in and habibul.pma17@iitp.ac.in}

\begin{document}

\maketitle

\begin{abstract}
In this article, we study the skew cyclic codes over $R_{k}=F_{p}+uF_{p}+\dots +u^{k-1}F_{p}$ of length $n$. We characterize the skew cyclic codes of length $n$ over $R_{k}$ as free left $R_{k}[x;\theta]$-submodules of $R_{k}[x;\theta]/\langle x^{n}-1\rangle$ and construct their generators and minimal generating sets. Also, an algorithm has been provided to encode and decode these skew cyclic codes.
\end{abstract}

{keywords}
Skew polynomial ring; Cyclic code, Skew cyclic code; Generating set; Syndrome decoding
{keywords}

\section{Introduction}

Many new error-correcting codes have been obtained from the cyclic codes due to advancement in algebraic structures. It is a very useful linear code and extensively studied in the theory of error-correcting codes for last few decades by many researchers. A cyclic code of length $n$ over a field $F$ is defined to be an ideal of the quotient ring $F[x]/\langle x^{n}-1\rangle$ and also this ideal is principally generated by a factor of $x^{n}-1$. But whenever we consider analogous concept of cyclic codes over finite ring, in general, we can not identify a cyclic code with a principally generated ideal of a quotient ring. Hence, task of finding generators of cyclic codes over finite ring is little bit difficult. In 2007, Abualrub and Siap \cite{Abualrub07} studied the cyclic codes over $\mathbb{Z}_{2}+u\mathbb{Z}_{2}$ and $\mathbb{Z}_{2}+\mathbb{Z}_{2}+u^{2}\mathbb{Z}_{2}$. They obtained the generators of cyclic codes over these two rings explicitly. Later on, in 2015, Singh and Kewat \cite{singh 15} generalized the concept of \cite{Abualrub07} over the rings $\mathbb{Z}_{p}[u]/<u^{k}>$ and discussed the generators and minimal spanning set of cyclic codes over it.\\
In 2007, Boucher et al. in \cite{Db07} used the concept of skew polynomial rings into the coding theory. They characterized skew cyclic code of length $n$ as an ideal of the quotient ring $F[x;\theta]/\langle x^{n}-1\rangle$ where $\theta$ is an automorphism on the field $F$. Later Siap et al. \cite{siap11}, consider the  skew cyclic codes of length $n$ as $F[x;\theta]$-submodules of $F[x;\theta]/\langle x^{n}-1\rangle$ and also they have constructed generators of these skew cyclic codes as the submodules. Recently, Dastbasteh et al. in \cite{Dastbast17}, studied the skew cyclic codes over the ring $F_{p}+uF_{p}$ and obtained their generators.\\
Let $F_{p}$ be the Galois field of $p$ elements. For any odd prime $p$ and integer $k\geq 1$, let $R_{k}=F_{p}+uF_{p}+\dots +u^{k-1}F_{p}$ where $u^{k}=0$. In this article, we study the skew cyclic codes of length $n$ over $R_{k}$ and construct their generators explicitly. The main motive of the study is to drive skew cyclic codes by using their generators and show how to encode and decode these skew cyclic codes over $R_{k}$. Note that the ring $R_{k}$ is isomorphism to the quotient ring $F_{p}[u]/\langle u^{k}\rangle$ and $R_{k-i}$ is subring of $R_{k}$ for any $k \geq i\geq 1$. Any element $w$ of the ring $R_{k}$ can be written as $w=a_{0}+ua_{1}+\dots +u^{k-1}a_{k-1}$ where $a_{i}\in F_{p}$. Let $\theta$ be an element of the Galois group Aut$(R_{k})$ and $\theta(u)=a_{0}+ua_{1}+\dots +u^{k-1}a_{k-1}$. Since $\theta$ is an automorphism on $R_{k}$ and $u^{k}=0$, so $\theta(u^{k})=0$ and hence $a_{0}=0$. Particularly, we choose the automorphism $\theta$ as $\theta(1)=1$ and $\theta(u)=su$, for some non-zero $s$ in $F_{p}$. For this automorphism $\theta$, the set $R_{k}[x;\theta]=\big \{ a_{0}+a_{1}x+\dots +a_{n}x^{n}\mid a_{i}\in R_{k} \big \}$ forms a non-commutative ring under addition of polynomials and multiplication of polynomials with respect to the condition $ax^{i} bx^{j}=a\theta^{i}(b)x^{i+j}$. This ring is known as skew polynomial ring. Let order of the automorphism $\theta$ is $m$, i.e. $\theta^{m}(a)=a$ for all $a\in R_{k}$. One can see that center of $R_{k}[x;\theta]$ is $F_{p}[x^{m}]$ and hence $R_{k, n}=R[x;\theta]/\langle x^{n}-1\rangle$ is a ring when $n\mid m$, where skew cyclic codes of length $n$ over $R_{k}$ are nothing but ideals of the quotient ring $R_{k, n}$. However, $R_{k, n}$ is a left $R_{k}[x;\theta]$-module and skew cyclic codes of length $n$ over $R_{k}$ are left $R_{k}[x;\theta]$-submodules of $R_{k, n}$. Since we are interested to get skew cyclic code of arbitrary length $n$ over $R_{k}$, we focus on the module structure of $R_{k, n}$ throughout this note.

\section{Definitions and Basic Results}
\df A code of length $n$ over $R_{k}$ is said to be skew cyclic code if \\
\begin{enumerate}
\item $C$ is a submodule of $R^{n}_{k}$;
\item For any $c=(c_{0}, c_{1},\dots c_{n-1})\in C$, we have $\tau(c)=(\theta(c_{n-1}), \theta(c_{0}),\dots ,\theta(c_{n-2}))\in C$.
\end{enumerate}
Note that the above definition is nothing but the definition of cyclic code over $R_{k}$ when $\theta$ is an identity automorphism. For any codeword $c=(c_{0}, c_{1},\dots c_{n-1})\in R^{n}_{k}$, we can find a polynomial $c(x)=c_{0}+c_{1}x+\dots c_{n-1}x^{n-1}$ in $R_{k, n}=R_{k}[x;\theta]/\langle x^{n}-1 \rangle.$ With this identification one can easily find the following result.\\
\theorem
A linear code $C$ of length $n$ over $R_{k}$ is a skew cyclic code if and only if the polynomial representation of $C$ is an $R_{k}[x;\theta]$-submodule of $R_{k, n}=R_{k}[x;\theta]/\langle x^{n}-1 \rangle.$

\theorem \cite{McDonald} \label{div}
Let $f(x), g(x)\in R_{k}[x;\theta]$ where leading coefficient of $g(x)$  is a unit. Then there exist two unique polynomials $q(x), r(x)\in R_{k}[x;\theta]$ such that\\
\begin{align*}
f(x)=q(x)g(x)+r(x),
\end{align*}
where $r(x)=0$ or $deg(g(x))>deg(r(x))$. \\
This theorem is known as right division algorithm. Similar result can be stated for left division.

\proposition \label{pro 1} For any polynomial $f(x)\in R_{k}[x;\theta]$, there exist polynomials $f_{i}(x)\in R_{k-i}[x;\theta]$ such that $f(x)u^{i}=u^{i}f_{i}(x)$ for $1\leq i\leq k-1$.
\begin{proof}
Let $f(x)\in R_{k}[x;\theta]$. Then $f(x)=\sum (a_{0, i}+ua_{1,i}+\dots u^{k-1}a_{k-1, i})x^{i} $ where $a_{j, i}\in F_{p}$, for $0\leq j\leq k-1$.\\\\
Now,
\begin{align} \label{equ 1}
\nonumber f(x)u&=\sum (a_{0, i}+ua_{1,i}+\dots u^{k-1}a_{k-1, i})x^{i}u\\
\nonumber &=\sum (a_{0, i}\theta^{i}(u)+ua_{1,i}\theta^{i}(u)+\dots u^{k-1}a_{k-1, i}\theta^{i}(u))x^{i}\\
\nonumber &=\sum (a_{0, i}s^{i}u+ua_{1,i}s^{i}u+\dots u^{k-1}a_{k-1, i}s^{i}u)x^{i}\\
\nonumber &=\sum (a_{0, i}s^{i}u+ua_{1,i}s^{i}u+\dots u^{k-2}a_{k-2, i}s^{i}u)x^{i}\\
&=uf_{1}(x)
\end{align}
where $f_{1}(x)=\sum (a_{0, i}s^{i}+ua_{1,i}s^{i}+\dots u^{k-2}a_{k-2, i}s^{i})x^{i}\in R_{k-1}[x;\theta]$.  Again, multiplying $u$ from right side of the equation (\ref{equ 1}), we get \\
\begin{align*}
f(x)u^{2}&=u\sum (a_{0, i}s^{2i}u+ua_{1,i}s^{2i}u+\dots u^{k-2}a_{k-2, i}s^{2i}u)x^{i}\\
&=u^{2}\sum (a_{0, i}s^{2i}+ua_{1,i}s^{2i}+\dots u^{k-3}a_{k-3, i}s^{2i})x^{i}\\
&=u^{2}f_{2}(x)
\end{align*}
where $f_{2}(x)=\sum (a_{0, i}s^{2i}+ua_{1,i}s^{2i}+\dots u^{k-3}a_{k-3, i}s^{2i})x^{i}\in R_{k-2}[x;\theta]$. Continuing this process, we get $f(x)u^{3}=u^{3}f_{3}(x),\dots , f(x)u^{k-1}=u^{k-1}f_{k-1}(x)$ where each $f_{i}(x)\in R_{k-i}[x;\theta]$.
\end{proof}

\proposition \label{pro 2} The set of units of $R_{i}[x;\theta]$ is $U(R_{i} [x;\theta])=\big \{ a+uh_{1}(x)+u^{2}h_{2}(x)+\dots +u^{i-1}h_{i-1}(x) \mid a\in F^{*}_{p}$ and $h_{i}(x)\in F_{p}[x]\big \}$.

\begin{proof}
We proof this result by induction on $i$. For $i=1$, $R_{1}=F_{p}$ and hence $U(F_{p}[x])=F^{*}_{p}$. For $i=2$, Lemma 13 of \cite{Dastbast17} gives the result. Assume that the result is true for $i=m>2$. In order to prove the result for $i=m+1$, let $a+uh_{1}(x)+\dots u^{m}h_{m}(x) \in R_{m+1}[x;\theta]$ where $a\in F^{*}_{p}$ and $h_{i}(x)\in F_{p}[x]$. Then $g(x)=a+uh_{1}(x)+\dots u^{m-1}h_{m-1}(x)\in U(R_{m}[x;\theta])$.\\
Now,
\begin{align*}
(g+u^{m}h_{m})(g^{-1}-g^{-1}u^{m}h_{m}g)&=1-u^{m}h_{m}g+u^{m}h_{m}g-u^{m}h_{m}g^{-1}u^{m}h_{m}g\\
&=1-u^{2m}h_{t}h_{m}g ~( ~By ~~Proposition ~~(\ref{pro 1}))\\
&=1 ~(~as~~ 2m>m+1, u^{2m}=0).
\end{align*}
Hence, $a+uh_{1}(x)+\dots u^{m}h_{m}(x) \in U(R_{m+1}[x;\theta])$.\\
Conversely, let $f(x)\in R_{m+1}[x;\theta]$ be a unit. Then there exists a polynomial $g(x)$ in $R_{m+1}[x;\theta]$ such that $f(x)g(x)=g(x)f(x)=1$. Therefore, $f_{0}(x)g_{0}(x)=1$ where $f(x)=f_{0}(x)+uf_{1}(x)+\dots +u^{m}f_{m}(x)$ and $g(x)=g_{0}(x)+ug_{1}(x)+\dots +u^{m}g_{m}(x)$. This shows that $f_{0}(x)$ is a non-zero constant polynomial in $F_{p}[x]$. Hence, $f(x)=a+uf_{1}(x)+\dots +u^{m}f_{m}(x)$ where $a\in F^{*}_{p}.$ This completes the proof.
\end{proof}

\proposition \label{pro 3} For $k\geq 2$, the polynomial $x^{n}-1$ factories in $R_{k-1}[x;\theta]$ as $x^{n}-1=g_{1}(x)g_{2}(x)$ if and only if there exist polynomials $f_{1}(x), f_{2}(x)\in R_{k}[x;\theta]$ such that $x^{n}-1=f_{1}(x)f_{2}(x)$ where $f_{i}(x)=g_{i}(x)+u^{k-1}k_{i}(x)$, $k_{i}(x)\in F_{p}(x)$.
\begin{proof}
Let $x^{n}-1=f_{1}(x)f_{2}(x)$ in $R_{k}[x;\theta]$ where $f_{i}(x)=g_{i}(x)+u^{k-1}k_{i}(x)$, $g_{i}(x)\in R_{k-1}[x;\theta]$.\\ Then
\begin{align*}
x^{n}-1&=(g_{1}(x)+u^{k-1}k_{1}(x))(g_{2}(x)+u^{k-1}k_{2}(x))\\
&=g_{1}(x)g_{2}(x)+g_{1}(x)u^{k-1}k_{2}(x)+u^{k-1}k_{1}(x)g_{2}(x)+u^{k-1}k_{1}(x)u^{k-1}k_{2}(x)\\
&=g_{1}(x)g_{2}(x)+u^{k-1}g'_{1}(x)k_{2}(x)+u^{k-1}k_{1}(x)g_{2}(x)+u^{2k-2}k'_{1}(x)k_{2}(x) \\
&\hspace{.5cm}~(~By ~Proposition~ (\ref{pro 1}))
\end{align*}
Therefore, in $R_{k-1}[x;\theta]$, we have $x^{n}-1=g_{1}(x)g_{2}(x)$.\\
Conversely, as $R_{k-1}[x;\theta]$ is a subring of the ring $R_{k}[x;\theta]$, so factorization $x^{n}-1=g_{1}(x)g_{2}(x)$ in $R_{k-1}[x;\theta]$ can be treated in $R_{k}[x;\theta]$ as well. In fact, in this case $f_{i}(x)=g_{i}(x)$ and $k_{i}(x)=0$.
\end{proof}

\cor \label{cor 1} The polynomial $x^{n}-1$ factories in $F_{p}[x]$ as $x^{n}-1=g_{1}(x)g_{2}(x)$ if and only if there exist polynomials $f_{1}(x), f_{2}(x)\in R_{k}[x;\theta]$ such that $x^{n}-1=f_{1}(x)f_{2}(x)$ where $f_{i}(x)=g_{i}(x)+ul_{1 i}(x)+u^{2}l_{2 i}+\dots +u^{k-1}l_{k-1 i}$, $l_{j i}\in F_{p}[x].$
\begin{proof}
Repeated application of Proposition (\ref{pro 3}).
\end{proof}

\section{Generators of skew cyclic codes over $R_{k}$}
In this section, we are interested to find the generators of skew cyclic codes of arbitrary length $n$ over $R_{k}$. Note that if $f(x)\in R_{k}[x;\theta]$ and leading coefficient $d$ of $f(x)$ is unit, then $d^{-1}f(x)$ is a monic polynomial. But by Proposition \ref{pro 2}, every element of $R_{k}$ is not a unit. Therefore, a skew cyclic code $C$ over $R_{k}$ may or may not contains any monic polynomials. By simple application of Theorem \ref{div}, we can find the generators if $C$ contains any monic polynomial of minimal degree. However, task would be difficult if $C$ does not contains any monic polynomial or contains some monic polynomials which are not of minimal degree in $C$. Based on this distinction, we would be able to find generators of skew cyclic code for all possible cases in next four theorems.

\theorem Let $f(x)$ be a non-monic polynomial of minimal degree in $C$. Then $f(x)=u^{i}a^{i}(x)$ where $a^{i}(x)\in R_{k-i}[x;\theta]$ for some positive integer $i$.
\begin{proof}
Let $f(x)$ be a non-monic polynomial of minimal degree in $C$. Let $f(x)=a_{0}+a_{1}x+\dots +ua_{r}x^{r}$ where $a_{i}\in R_{k}$ and $a_{r}\in R_{k-1}$. Suppose $a_{r}=t_{1}+ut_{2}+\dots +u^{k-2}t_{k-1}$  where $t_{i}\in F_{p}$. Let $i$ be a least positive integer such that $t_{i}\neq 0$. Then $f(x)=a_{0}+a_{1}x+\dots +ua_{r}x^{r}$ with $a_{r}=u^{i-1}(t_{i}+ut_{i+1}+\dots +u^{k-i-1}t_{k-i})$. Therefore, $u^{k-1}f(x), u^{k-2}f(x), \dots, u^{k-i}f(x)$ all are in $C$ with degree less than $r$. Since $f(x)$ is the minimal polynomial in $C$ with degree $r$, $u^{k-1}f(x)=0, u^{k-2}f(x)=0, \dots ,u^{k-i}f(x)=0$. Hence, $f(x)=u^{i}a^{i}(x)$ where $a^{i}(x)\in R_{k-i}[x;\theta]$ with unit leading coefficient.
\end{proof}

\theorem \label{th 1} Let $C$ be a non-zero skew cyclic codes of length $n$ over $R_{k}$. Suppose $C$ does not contain any monic polynomial and $a(x)=ua^{1}(x)$ is the minimal degree polynomial in $C$, then $C=\langle ua^{1}(x)\rangle$ with $x^{n}-1=b^{1}(x)a^{1}(x)$ in $R_{k-1}[x;\theta]$.

\begin{proof}
Let $c(x)\in C$ be a codeword. Then $c(x)$ is not a monic polynomial. Let $c(x)=c_{0}+c_{1}x+\dots +uc_{t}x^{t}$ where $c_{i}\in R_{k}$ and $c_{t}\in R_{k-1}$. Assume that $c_{t}=e_{1}+ue_{2}+\dots +u^{k-2}e_{k-1}\in R_{k-1}$. Let $j$ be a least positive integer such that $e_{j}\neq 0$. Then $c(x)=c_{0}+c_{1}x+\dots +u^{j}c_{t}x^{t}$ where $c_{i}\in R_{k}$ and $c_{t}\in R^{*}_{k-j}$. Let $a(x)=ua^{1}(x)$ be the polynomial of minimal degree $r$ in $C$ with $a^{1}(x)=a_{0}+a_{1}x+\dots +a_{r}x^{r}$ where $a_{i}\in R_{k-1}$ and $a_{r}\in R^{*}_{k-1}$. Let $c(x)=c_{1}(x)+c_{2}(x)$ where $c_{1}(x)$ contains all terms of degree up to $r-1$ and $c_{2}(x)$ contains all rest terms (with degree $r$ and above). If possible, let $c_{t-1}$ be a unit. Since $a_{r}\in R^{*}_{k-1}$, $\alpha=\theta(a_{r})\in R^{*}_{k-1}$. Also, by definition, $\theta^{i}(u)=s^{i}u$ and let $d=s^{t-r}\in F^{*}_{p}$. Now,
\begin{align*}
d_{1}(x)&=u^{j-1}d^{-1}\alpha^{-1}x^{t-r}a(x)\\
&=u^{j-1}d^{-1}\alpha^{-1}\theta(ua_{0})x^{t-r}+u^{j-1}d^{-1}\alpha^{-1}\theta(ua_{1})x^{t-r+1}+\dots u^{j-1}d^{-1}\alpha^{-1}\theta(ua_{r-1})x^{t-1}\\
&\hspace{.5cm} +u^{j-1}\alpha^{-1}d^{-1}\theta(ua_{r})x^{t}\\
&=u^{j-1}d^{-1}\alpha^{-1}\theta(ua_{0})x^{t-r}+u^{j-1}d^{-1}\alpha^{-1}\theta(ua_{1})x^{t-r+1}+\dots u^{j-1}d^{-1}\alpha^{-1}\theta(ua_{r-1})x^{t-1}\\
&\hspace{.5cm} +u^{j}x^{t}\in C.
\end{align*}
Also,
\begin{align*}
d_{2}(x)&=(c_{t})^{-1}c(x)\\
&=(c_{t})^{-1}c_{0}+(c_{t})^{-1}c_{1}x+\dots +(c_{t})^{-1}c_{t-1}x^{t-1}+u^{j}x^{t}\in C.
\end{align*}
Then $d(x)=d_{2}(x)-d_{1}(x)\in C$ is a polynomial of degree $(t-1)$. Hence, by Proposition \ref{pro 2}, the leading coefficient of $d(x)$ i.e.,$d_{t-1}$ is a unit. This contradicts the fact that $C$ does not contain any monic polynomial. Thus, $c_{t-1}$ can not be unit. With similar argument, we can conclude that none of the coefficients of $c_{2}(x)$ is a unit. \\If possible, let coefficient $c_{i}$ of the polynomial $c_{1}(x)$ be a unit for some integer $i$. Then $u^{k-1}c(x)=u^{k-1}c_{1}(x)\in C$ where $deg(u^{k-1}c_{1}(x))<r=$deg$(a(x)).$ This contradicts the fact that $a(x)$ is the minimal degree polynomial in $C$. Consequently, $c(x)=uc^{1}(x)$ where $c^{1}(x)\in R_{k-1}[x;\theta]$.\\
Again, by division algorithm, we get two polynomials $b^{1}(x), r^{1}(x)$ in $R_{k-1}[x;\theta]$ such that\\
\begin{align*}
x^{n}-1=b^{1}(x)a^{1}(x)+r^{1}(x),
\end{align*}
where $deg(r^{1}(x))< deg(a^{1}(x))$ or $r^{1}(x)=0$. By Proposition \ref{pro 1}, we have \\
\begin{align*}
u(x^{n}-1)=b^{1}_{1}(x)ua^{1}(x)+ur^{1}(x).
\end{align*}
Thus, in $R_{n, k}=R_{k}[x;\theta]/\langle x^{n}-1\rangle$, we have $ur^{1}(x)=-b^{1}_{1}(x)ua^{1}(x)\in C$. As $deg(ur^{1}(x))=$deg$(r^{1}(x))<deg(a^{1}(x))=r$, $ur^{1}(x)=0$. Since $r^{1}(x)\in R_{k-1}[x;\theta]$, $r^{1}(x)=0$. Therefore, $x^{n}-1=b^{1}(x)a^{1}(x)$ in $R_{k-1}[x;\theta]$.
\end{proof}

\theorem \label{th 2} Let $C$ be a non-zero skew cyclic code of length $n$ over $R_{k}$ in which  $g(x)$ be the minimal degree monic polynomial. Then $C=\langle g(x)\rangle$ and $x^{n}-1=k(x)g(x)$ in $R_{k, n}$.

\begin{proof}
One can proof it by simple application of Theorem \ref{div}.
\end{proof}

\theorem \label{th 3} Let $C$ be a non-zero skew cyclic code of length $n$ with at least one monic polynomial over $R_{k}$. Suppose $C$ does not contain any monic polynomial of minimal degree and $a(x)=ua^{1}(x)$ is the polynomial of  minimal degree in $C$. Let $g(x)$ be the polynomial of minimal degree among the monic polynomials in $C$. Then $C=\langle g(x)+up_{1}(x)+\dots + u^{k-1}p_{k-1}(x), ua^{1}(x)\rangle$ where $x^{n}-1=k(x)g(x)$ in $R_{k}[x;\theta]$ and $x^{n}-1=b^{1}(x)a^{1}(x)$ in $R_{k-1}[x;\theta]$.
\begin{proof}
Let $f(x)$ be the polynomial of minimal degree among the monic polynomials in $C$ and $a(x)=ua^{1}(x)$ be the polynomial of minimal degree in $C$ which is not monic. Let $c(x)$ be a codeword in $C$. By division algorithm, there exist two polynomials $q_{1}(x), r_{1}(x)$ such that \\
\begin{align*}
c(x)=q_{1}(x)f(x)+r_{1}(x),
\end{align*}
where $deg(f(x))> deg(r_{1}(x))$ or $r_{1}(x)=0$. Therefore, $r_{1}(x)\in C$ and hence $r_{1}(x)$ is non-monic. By the proof of Theorem \ref{th 1}, we have $r_{1}(x)=ur_{2}(x)$. Again, by division algorithm, we have\\
\begin{align*}
r_{2}(x)=q_{2}(x)a^{1}(x)+r_{3}(x),
\end{align*}
where $deg(a^{1}(x)) > deg(r_{3}(x))$ or $r_{3}(x)=0$. Therefore, $r_{1}(x)=ur_{2}(x)=q'_{2}(x)ua^{1}(x)+ur_{3}(x)$, and this implies $ur_{3}(x)\in C$. Since, $deg(a^{1}(x)) > deg(r_{3}(x)) = deg(ur_{3}(x))$, so $ur_{3}(x)=0$. As $r_{3}(x)\in R_{k-1}[x;\theta]$, $r_{3}(x)=0$. Hence, \\
\begin{align*}
c(x)&=q_{1}(x)f(x)+r_{1}(x)\\
&=q_{1}(x)f(x)+ur_{2}(x)\\
&=q_{1}(x)f(x)+q'_{2}(x)ua^{1}(x)\in \langle f(x), ua^{1}(x) \rangle.
\end{align*}
Consequently, $C=\langle f(x), ua^{1}(x) \rangle.$
By following the proof of Theorem \ref{th 1}, we can conclude $x^{n}-1=b^{1}(x)a^{1}(x)$ in $R_{k-1}[x;\theta]$.\\
Further, let $f(x)=f_{1}(x)+uf_{2}(x)+\dots u^{k-1}f_{k}(x)$ where $f_{i}(x)\in F_{p}[x]$. Then, by division algorithm, we have\\
\begin{align*}
x^{n}-1=q_{4}(x)(f_{1}(x)+uf_{2}(x)+\dots u^{k-1}f_{k}(x))+ r_{4}(x),
\end{align*}
where $r_{4}(x)=0$ or $deg(r_{4}(x))< deg(f_{1}(x))= deg(f(x))$. Therefore, $r_{4}(x)\in C$ and this shows that $r_{4}(x)$ is not a monic polynomial. Hence, from the proof of Theorem \ref{th 1}, we conclude that $r_{4}(x)=l(x)ur_{5}(x)=ul'(x)r_{5}(x)$. Thus,\\
\begin{align*}
x^{n}-1=q_{4}(x)(f_{1}(x)+uf_{2}(x)+\dots u^{k-1}f_{k}(x))+ ul'(x)r_{5}(x).
\end{align*}
In $F_{p}[x]$, we have \\
\begin{align*}
x^{n}-1=q_{6}(x)f_{1}(x).
\end{align*}
Now, by Corollary \ref{cor 1}, there exists a polynomial $g(x)$ in $R_{k}[x;\theta]$ which is a right divisor of $x^{n}-1$ in $R_{k}[x;\theta]$ such that $g(x)=f_{1}(x)+ue_{1}(x)+\dots u^{k-1}e_{k-1}(x)$ with $deg(g(x)) = deg(f_{1}(x))$. Therefore, \\
\begin{align*}
f(x)&=f_{1}(x)+uf_{2}(x)+\dots u^{k-1}f_{k}(x)\\
&=g(x)+u(f_{2}(x)-e_{1}(x))+\dots u^{k-1}(f_{k}(x)-e_{k-1}(x)).
\end{align*}
Consequently, $C=\langle f(x), ua^{i}(x)\rangle =\langle g(x)+up_{1}(x)+\dots +u^{k-1}p_{k-1}(x), ua^{1}(x)\rangle$ where $x^{n}-1 = k(x)g(x)$ in $R_{k}[x;\theta]$ and $x^{n}-1=b^{1}(x)a^{1}(x)$ in $R_{k-1}[x;\theta]$.
\end{proof}

\proposition \label{pro 4} Let $C=\langle g(x)+up_{1}(x)+\dots +u^{k-1}p_{k-1}(x), ua^{1}(x)\rangle$ be a non-zero skew cyclic code as given in Theorem \ref{th 3}. Then $a^{1}(x)\mid (g(x)+up_{1}(x)+\dots +u^{k-2}p_{k-2}(x))$ mod $u^{k-1}$ and $\frac{x^{n}-1}{g(x)}[up_{1}(x)+\dots +u^{k-1}p_{k-1}(x)]\in \langle ua^{1}(x)\rangle$.
\begin{proof}
Let $ut(x)\in C=\langle g(x)+up_{1}+\dots +u^{k-1}p_{k-1}(x), ua^{1}(x)\rangle$ where $t(x)\in R_{k-1}[x;\theta]$. By division algorithm, we have\\
\begin{align*}
t(x)=q(x)a^{1}(x)+r(x),
\end{align*}
where $deg(r(x)) < deg(a^{1}(x))$ or $r(x)=0$. Then $ut(x)=q'(x)ua^{1}(x)+ur(x)$ and hence $ur(x)\in C$. This is a contradiction unless $ur(x)=0, i.e., r(x)=0$. Therefore, $ut(x)=q'(x)ua^{1}(x)\in \langle ua^{1}(x)\rangle$. As the result, for any $ut(x)\in C$, we get $t(x)=q(x)a^{1}(x)$ and $ut(x)\in \langle ua^{1}(x)\rangle$. Note that $u( g(x)+up_{1}(x)+\dots +u^{k-1}p_{k-1}(x) )= u(g(x)+up_{1}(x)+\dots +u^{k-2}p_{k-2}(x))\in C$. Hence, from the above discussion we conclude that $a^{1}(x)\mid (g(x)+up_{1}(x)+\dots +u^{k-2}p_{k-2}(x))$ mod $u^{k-1}$. Further,\\
\begin{align*}
&\frac{x^{n}-1}{g(x)}[g(x)+up_{1}(x)+\dots +u^{k-1}p_{k-1}(x)]\\
&=u[(\frac{x^{n}-1}{g(x)})'p_{1}(x)+\dots +(\frac{x^{n}-1}{g(x)})'u^{k-2}p_{k-1}(x)]\in C.
\end{align*}
Thus, $\frac{x^{n}-1}{g(x)}[up_{1}(x)+\dots +u^{k-1}p_{k-1}(x)]\in \langle ua^{1}(x)\rangle$.
\end{proof}

4\section{Minimal spanning set}
In this section, we discuss the minimal spanning set of skew cyclic codes of length $n$ for different cases as given in Theorem \ref{th 1}, \ref{th 2} and \ref{th 3}. These minimal spanning sets will help to find generator matrices and cardinality of the skew cyclic codes over $R_{k}.$
\theorem \label{min th 1} Let $C=\langle ua^{1}(x)\rangle$ be a non-zero skew cyclic code of length $n$ over $R_{k}$ where $a(x)=ua^{1}(x)$ be the polynomial of minimal degree $r$ in $C$ with $x^{n}-1=b^{1}(x)a^{1}(x)$ in $R_{k-1}[x;\theta]$. Then \\
\begin{align*}
\Gamma= \big \{ ua^{1}(x), xa^{1}(x), \dots, x^{n-r-1}ua^{1}(x)\big \}
\end{align*}
forms a minimal generating set for the code $C$ and $\mid C\mid =(p^{k-1})^{n-r}.$
\begin{proof}
Let $C=\langle ua^{1}(x)\rangle$ where $x^{n}-1=b^{1}(x)a^{1}(x)$ in $R_{k-1}[x;\theta]$. Let $c(x)\in C$. Then $c(x)=j(x)ua^{1}(x)$. Let $j(x)=j_{1}(x)+u^{k-1}j_{2}(x)$ where $j_{1}(x)\in R_{k-1}[x;\theta]$. Then $c(x)=j(x)ua^{1}(x)=(j_{1}(x)+u^{k-1}j_{2}(x))ua^{1}(x)=uj_{3}(x)a^{1}(x)$. If $deg(j_{3}(x)) \leq (n-r-1)$, then $c(x)\in$ span$(\Gamma)$. Otherwise, by division algorithm, we have\\
\begin{align*}
j_{3}(x)=q_{1}(x)\frac{x^{n}-1}{a^{1}(x)}+r_{1}(x),
\end{align*}
where $deg(r_{1}(x)) < deg(a^{1}(x))=(n-r)$ or $r_{1}(x)=0$.\\
Therefore, \\
\begin{align*}
c(x)=uj_{3}(x)a^{1}(x)&=u(q_{1}(x)\frac{x^{n}-1}{a^{1}(x)}+r_{1}(x))a^{1}(x)\\
&=ur_{1}(x)a^{1}(x).
\end{align*}
Since $deg(r_{1}(x))\leq (n-r-1)$, so $c(x)\in$ span$(\Gamma).$ Clearly none of the element of $\Gamma$ is a linear combination of preceding elements. Therefore, $\Gamma$ is the minimal generating set for the skew cyclic code $C$. Since $r_{1}(x)\in R_{k-1}$, so $\mid C\mid = (p^{k-1})^{n-r}.$
\end{proof}

\theorem \label{min th 2} Let $C=\langle g(x)\rangle$ be a non-zero skew cyclic code of length $n$ where $g(x)$ be the monic polynomial of minimal degree $r$ in $C$ and $x^{n}-1=k(x)g(x)$ in $R_{k, n}$. Then\\
\begin{align*}
\Gamma=\{ g(x), xg(x), \dots, x^{n-r-1}g(x) \}
\end{align*}
forms a minimal generating set for the code $C$ and $\mid C\mid = (p^{k})^{n-r}$.
\begin{proof}
Similar proof as Theroem \ref{min th 1}.
\end{proof}

\theorem \label{min th 3} Let $C=\langle g(x)+up_{1}(x)+\dots +u^{k-1}p_{k-1}(x), ua^{1}(x)\rangle$ be a non-zero skew cyclic code of length $n$ over $R_{k}$ where $a(x)=ua^{1}(x)$ is the polynomial of minimal degree $t$ in $C$ which is not monic, $g(x)$ is the polynomial of minimal degree $r$ among all monic polynomials in $C$, $x^{n}-1=k(x)g(x)$ in $R_{k}[x;\theta]$ and  $x^{n}-1=b^{1}(x)a^{1}(x)$ in $R_{k-1}[x;\theta]$. Then \\
\begin{align*}
\Gamma=\big \{ h(x), xh(x), \dots ,x^{n-r-1}h(x), ua^{1}(x), xa^{1}(x), \dots x^{r-t-1}a^{1}(x)\big \}
\end{align*}
forms a minimal generating set for the code $C$ and $\mid C\mid =(p^{k})^{n-r}(p^{k-1})^{r-t}$, where $h(x)=g(x)+up_{1}(x)+\dots +u^{k-1}p_{k-1}(x)$.
\begin{proof}
Let $c(x)\in C$. Then $c(x)=j_{1}(x)(g(x)+up_{1}(x)+\dots +u^{k-1}p_{k-1}(x))+j_{2}(x)ua^{1}(x).$ If $deg(j_{1}(x))\leq (n-r-1)$, then $j_{1}(x)(g(x)+up_{1}(x)+\dots +u^{k-1}p_{k-1}(x))\in span(\Gamma)$. Otherwise, by division algorithm, we have\\
\begin{align*}
j_{1}(x)=q_{1}(x)\frac{x^{n}-1}{g(x)}+r_{1}(x),
\end{align*}
where $r_{1}(x)=0$ or $deg(r_{1}(x))<deg(\frac{x^{n}-1}{g(x)})=n-r.$\\
Therefore,
\begin{align*}
c(x)&=j_{1}(x)(g(x)+up_{1}(x)+\dots +u^{k-1}p_{k-1}(x))+j_{2}(x)ua^{1}(x)\\
&=(q_{1}(x)\frac{x^{n}-1}{g(x)}+r_{1}(x))(g(x)+up_{1}(x)+\dots +u^{k-1}p_{k-1}(x))+j_{2}(x)ua^{1}(x)\\
&=u[q'_{1}(x)p_{1}(x)+j'_{2}(x)a^{1}(x)+uq_{2}(x)p_{2}(x)+\dots +u^{k-2}p_{k-1}(x)]+ \\
&\hspace{.5cm}r_{1}(x)(g(x)+up_{1}(x)+\dots +u^{k-1}p_{k-1}(x)).
\end{align*}
Since $deg(r_{1}(x))\leq (n-r-1)$, so $r_{1}(x)(g(x)+up_{1}(x)+\dots +u^{k-1}p_{k-1}(x))\in$ span$(\Gamma)$. Now, we prove $uk(x)\in$ span$(\Gamma)$ where $uk(x)\in C$.\\
For this, let $uk(x)\in C$ with $deg(k(x))\geq deg(g(x))$. Then by division algorithm, we have\\
\begin{align*}
k(x)=q(x)(g(x)+up_{1}(x)+\dots +u^{k-1}p_{k-1}(x))+r(x),
\end{align*}
where $r(x)=0$ or $deg(r(x))<$deg$(g(x)+up_{1}(x)+\dots +u^{k-1}p_{k-1}(x))=$deg$(g(x))=r$. Hence,\\
\begin{align*}
uk(x)=q'(x)u(g(x)+up_{1}(x)+\dots +u^{k-1}p_{k-1}(x))+ur(x).
\end{align*}
Now, $deg(q'(x))= deg(q(x))= deg(k(x))-r\leq n-r-1$. Therefore, $q'(x)u(g(x)+up_{1}(x)+\dots +u^{k-1}p_{k-1}(x))\in $span$(\Gamma).$
Finally, we prove $ur(x)\in $span$(\Gamma)$. Here, Note that $deg(ur(x))< deg(g(x))$ and $deg(ur(x)) \geq deg(a^{1}(x))$. Also, from the proof of Proposition \ref{pro 4}, we know that $ur(x)\in \langle ua^{1}(x)\rangle$, therefore,  $ur(x)=m(x)ua^{1}(x)=um'(x)a^{1}(x)$. Consequently, $ur(x)=l_{0}ua^{1}(x)+l_{1}xua^{1}(x)+\dots +l_{r-t-1}x^{r-t-1}ua^{1}(x)\in $span$(\Gamma)$. Since none of the element of the set $\Gamma$ is a linear combination of preceding elements, hence $\Gamma$ is a minimal generating set for the code $C$ and $\mid C\mid = (p^{k})^{n-r}(p^{k-1})^{r-t}$.
\end{proof}

\section{Encoding of the skew cyclic codes over $R_{k}$}
Now, we propose an encoding algorithm for skew cyclic codes of arbitrary length $n$ over $R$ as the application of Theorems \ref{min th 1}, \ref{min th 2} and \ref{min th 3}.

\theorem \label{enc th} Let $C$ be a skew cyclic code of length $n$ over $R_{k}$.\\ \\
\textbf{Case I:} If $C=\langle ua^{1}(x)\rangle$ where $a(x)=ua^{1}(x)$ is the polynomial of minimal degree $r$ in $C$ and $x^{n}-1=b^{1}(x)a^{1}(x)$ in $R_{k-1}[x;\theta]$, then any codeword $c(x)\in C$ can be encoded as\\
\begin{align*}
c(x)=[t_{0}(x)+ut_{1}(x)+\dots +u^{k-2}t_{k-2}(x)]ua^{1}(x),
\end{align*}
where $t_{0}(x)+ut_{1}(x)+\dots +u^{k-2}t_{k-2}(x)$ is a polynomial of degree $\leq (n-r-1)$ in $R_{k-1}[x;\theta].$\\ \\
\textbf{Case II:} If $C=\langle g(x)\rangle$ where $g(x)$ is the monic polynomial of minimal degree $r$ in $C$ and $x^{n}-1=k(x)g(x)$ in $R_{k, n}$, then any codeword $c(x)\in C$ can be encoded as\\
\begin{align*}
c(x)=[t_{0}(x)+ut_{1}(x)+\dots +u^{k-1}t_{k-1}(x)]g(x),
\end{align*}
where $t_{0}(x)+ut_{1}(x)+\dots +u^{k-1}t_{k-1}(x)$ is a polynomial of degree $\leq (n-r-1)$ in $R_{k}[x;\theta].$\\ \\
\textbf{Case III:} If $C=\langle g(x)+up_{1}(x)+\dots u^{k-1}p_{k-1}(x), ua^{1}(x) \rangle$ where $g(x)$ is the polynomial of minimal degree $r$ among all monic polynomials in $C$, $a(x)=ua^{1}(x)$ is the polynomial of minimal degree $\tau$ in $C$, $x^{n}-1=b^{1}(x)a^{1}(x)$ in $R_{k-1}[x;\theta]$, $x^{n}-1=k(x)g(x)$ in $R_{k}[x;\theta]$, $a^{1}(x)\mid (g(x)+up_{1}(x)+\dots +u^{k-2}p_{k-2}(x))$ mod $u^{k-1}$ and $\frac{x^{n}-1}{g(x)}[up_{1}(x)+\dots +u^{k-1}p_{k-1}(x)]\in \langle ua^{1}(x)\rangle$, then any codeword $c(x)\in C$ can be encoded as\\
\begin{align*}
c(x)&=[t_{0}(x)+ut_{1}(x)+\dots +u^{k-1}t_{k-1}(x)](g(x)+up_{1}(x)+\dots u^{k-1}p_{k-1}(x))\\
&\hspace{.5cm}+[j_{0}(x)+uj_{1}(x)+\dots +u^{k-2}j_{k-2}(x)]ua^{1}(x),
\end{align*}
where $t_{0}(x)+ut_{1}(x)+\dots +u^{k-1}t_{k-1}(x)$ is a polynomial of degree $\leq (n-r-1)$ in $R_{k}[x;\theta]$ and $j_{0}(x)+uj_{1}(x)+\dots +u^{k-2}j_{k-2}(x)$ is a polynomial of degree $\leq (r-\tau-1)$ in $R_{k-1}[x;\theta].$\\

Note that if the skew cyclic code $C$ is in the form given in Theorem \ref{th 3}, we follow the encoding process of Case III of Theorem \ref{enc th} and syndrome decoding process for decoding. Through an example we present the verification of encoding and decoding algorithm for Case III of Theorem \ref{enc th}.

\example Let $k=4$ and $p=3$, $R_{3} = F_{3}+uF_{3}+u^{2}F_{3}, R_{2} = F_{3}+uF_{3}$ and automorphism $\theta(u) = -u$ on $R_{3}$. Take $n = 6, r = 4, \tau = 2$ and $C=\langle x^{4}+(1+u+u^{2})x^{2}-(u+u^{2})x+(1+u+u^{2}), u(x^{2}-x+1)\rangle$. Suppose the sender wishes to transmit two strings $I = (1+u+2u^{2}, 2u+u^{2})\in R^{2}_{3}$ and $J = (2+u, u)\in R^{2}_{2}$. Following the Case III of Theorem \ref{enc th}, the sender encoded two strings $I, J$ as\\
\begin{align*}
Encode(I, J) &= ((1+u+2u^{2})x+2u+u^{2})(x^{4}+(1+u+u^{2})x^{2}-(u+u^{2})x+\\
&\hspace{.5cm}(1+u+u^{2}))+u((2+u)x+u)(x^{2}-x+1)\\
& = (1+u+2u^{2})x^{5}+(2u+u^{2})x^{4}+(1+2u)x^{3}+ux^{2}+(1+2u)x\\
&\hspace{.5cm}+2u+u^{2}.
\end{align*}
Therefore, the sender sends the encoded string $(1+u+2u^{2}, 2u+u^{2}, 1+2u, u, 1+2u, 2u+u^{2})$ though an open channel. Due to noise of the open channel, suppose receiver received the string as $(1+u+2u^{2}, 2u+2u^{2}, 1+2u, u, 1+2u, 2u+u^{2})$ (messages with some errors). Note that number of symbols in input is $10$ whereas in output is $18$. Also, the receiver follows the syndrome decoding algorithm to retrieved the actual string (messages) which was sent by sender. Thus, receiver should follow the following process.\\
\begin{align*}
t_{1}(x)+ut_{2}(x)+u^{2}t_{3}(x) &= (1+u+2u^{2})x^{5}+(2u+2u^{2})x^{4}+(1+2u)x^{3}+ux^{2}\\
&\hspace{.5cm}+(1+2u)x+2u+u^{2}. \\
\end{align*}
which gives $t_{1}(x) = x^{5}+x^{3}+x$, $t_{2}(x)= x^{5}+2x^{4}+2x^{3}+x^{2}+2x+2$ and $t_{3}(x)= 2x^{5}+2x^{4}+1.$ Therefore, syndromes are given by \\
\begin{align*}
e_{1}(x) &= t_{1}(x)(x^{2}-1)=0;\\
e_{2}(x) &= t_{2}(x)(x^{4}+x^{3}-x-1) = 0;\\
e_{3}(x) &= t_{3}(x)(x^{4}+x^{3}-x-1);\\
&= x+x^{2}-x-x^{5}.
\end{align*}
Moreover, \\
\begin{align*}
x^{4}(x^{4}+x^{3}-x-1) = x+x^{2}-x-x^{5}.
\end{align*}
Thus, $e_{3}(x)$ is the syndrome of $x^{4}$. Consequently, the receiver can detect the error term $u^{2}x^{4}$. Apply division algorithm, receiver obtained the strings $I$ and $J$ as follows:\\
\begin{eqnarray}\label{equation}
\nonumber &&(1+u+2u^{2})x^{5}+(2u+u^{2})x^{4}+(1+2u)x^{3}+ux^{2}+(1+2u)x+2u+u^{2}\\
\nonumber &&= ((1+u+2u^{2})x+2u+u^{2})(x^{4}+(1+u+u^{2})x^{2}-(u+u^{2})x+(1+u+u^{2}))\\
\nonumber &&\hspace{1.4cm}+u((2+u)x+u)(x^{2}-x+1).\\
\end{eqnarray}
Hence, receiver can extract $I$ and $J$ from above \ref{equation}.\\

In next example, we construct some skew cyclic codes of arbitrary length $n$ over $R$ as proposed in Theorem \ref{th 1} and \ref{th 2}.\\

\example
Let $F_{5}$ be the Galois field of order $5$ and $R_{3}=F_{5}+uF_{5}+u^{2}F_{5}$. Consider the automorphism $\theta$ on $R_{3}$ as $\theta(u)=-u$, i.e. $\theta(a+ub+u^{2}c)= a-ub+u^{2}c$ where $a, b, c\in R_{3}$.\\
Here, we are interested to show some principally generated skew cyclic codes of length $4$. First, we discuss skew cyclic codes with non-monic generators.\\
We consider one of the factorization of $x^{4}-1$ as $x^{4}-1=(ax^{2}+b)(cx^{2}+d)$ where $a,b,c,d\in F_{5}+uF_{5}$. This gives the possible factorization as given in Table 1.\\ \\
\begin{tabular}{ |p{5.50cm}|p{1.25cm}|p{2.50cm}|p{1.5cm}|p{1.5cm}|  }

 \hline
\multicolumn{5}{|c|}{\textbf{Table 1}: Principally generated skew cyclic codes of length $4$ over $R_{3}$} \\
 \hline
 $x^{4}-1= f_{1}(x)f_{2}(x)$ & No. of distinct factors &Codes generated by non-monic poly.& Rank(C)& Distance, d(C)\\
 \hline
$[(1+uk)x^{2}+1+uk][(1-uk)x^{2}+4+uk]$ & 10  & $C_{1}=\langle uf_{1}\rangle, C_{2}=\langle uf_{2}\rangle$ &  2&2\\
 \hline
$[(4-uk)x^{2}+1+uk][(4+uk)x^{2}+4+uk]$ &   10  & $C_{1}=\langle uf_{1}\rangle, C_{2}=\langle uf_{2}\rangle$   &2&2\\
 \hline
$[(2+uk)x^{2}+2+uk][(3+uk)x^{2}+2-uk]$ &10 & $C_{1}=\langle uf_{1}\rangle, C_{2}=\langle uf_{2}\rangle$&  2&2\\
 \hline
$[(2-uk)x^{2}+3+uk][(3-uk)x^{2}+3-uk]$   &10 & $C_{1}=\langle uf_{1}\rangle, C_{2}=\langle uf_{2}\rangle$&  2&2\\
 \hline

\end{tabular}\\ \\
Note that in the first row of Table 1, we factorized $x^{4}-1$ as \\
\begin{align*}
x^{4}-1= [(1+uk)x^{2}+4][(1-uk)x^{2}+1],
\end{align*}
where $k = 0, 1, 2, 3, 4.$ By putting $k=0$, we get
\begin{align*}
x^{4}-1&=(x^{2}+4)(x^{2}+1)\\
&=(x^{2}-1)(x^{2}+1).
\end{align*}
Moreover, $(x^{2}+1)$ can be factorized as $(x^{2}+1)=(ax+b)(cx+d)$ in $F_{5}+uF_{5}$ as follows:
\begin{enumerate}
\item $(x^{2}+1)=[(1+ut)x+2+us][(1+ut)x+3+us];$
\item $(x^{2}+1)=[(4+ut)x+2+us][(4+ut)x+3+us];$
\item $(x^{2}+1)=[(2+ut)x+1-us][(3+u(5-t))x+1+us];$
\item $(x^{2}+1)=[(2+ut)x+4-us][(3+u(5-t))x+4+us];$
\end{enumerate}
where $t, s = 0, 1, 2, 3, 4.$ Note that there are $4\times 50=200$ distinct linear factors in $R_{3}$. Therefore, there are 200 skew cyclic codes of length 4 over $R_{3}$ in which each has rank 3. Similarly, we can factorize $x^{2}-1$ in $(F_{5}+uF_{5})[x;\theta]$ to get more skew cyclic codes generated by non-monic polynomials over $R_{3}$. \\
Now, consider the skew cyclic codes over $R_{3}$ whose generators are monic polynomials in $C$ as given in Theorem \ref{th 2}.\\
Let $C = \langle f(x)\rangle = \langle (1+4u+u^{2})x^{2}+4+u+4u^{2}\rangle$ where $x^{n}-1=k(x)f(x)$ in $R_{3, n}$. Then the rank of the skew cyclic code $C$ is 2 while generators matrix $G$ and parity check matrix $H$ are given by\\
\[
G=
\begin{bmatrix}
    4+u+4u^{2}       & 0 & 1+4u+u^{2} & 0 \\
    o       & 4+u+4u^{2} & 0 & 1+4u+u^{2} \\
\end{bmatrix}
\]
and
\[
H=
\begin{bmatrix}
    1+u & 0 & 1+u & 0 \\
    0 & 1+u & 0 & 1+u \\
\end{bmatrix}.
\]

\section{Conclusion}
In this article, we study skew cyclic codes of arbitrary length $n$ over $R_{k}=F_{p}+uF_{p}+\dots +u^{k-1}F_{p}$ with $u^{k}=0$. The generators and minimal spanning sets of the skew cyclic codes over $R_{k}$ are obtained. Further, we proposed an algorithm to encode such skew cyclic codes.

\section*{Acknowledgement}
The authors are thankful to University Grant Commission(UGC), Govt. of India for financial support under Ref. No. 20/12/2015(ii)EU-V dated 31/08/2016 and Indian Institute of Technology Patna for providing the research facilities.

\end{document}